\documentclass{ws-procs975x65}

\begin{document}

~~~~~~~~~~~~~~~~~~~~~~~~~~~~~~~~~~~~~~~~~~~~~~~~~~~~~~~~~~~~
~~~~~~~~~~~~~~~~~~~~~~~~~~~~~~                                  IPMU 10-0047

\title{BLACK HOLES CONSTITUTE ALL DARK MATTER}

\author{Paul H. Frampton$^*$}

\address{Department of Physics and Astronomy,
University of North Carolina, Chapel Hill, USA {\it and}
Institute for the Mathematics and Physics of the Universe, University of Tokyo,
Japan.\\
$^*$E-mail: frampton@physics.unc.edu {\it and} 
paul.frampton@ipmu.jp}

\begin{abstract}The dimensionless entropy , ${\cal S} \equiv S/k$,
of the visible universe, taken as a sphere
of radius 50 billion light years with the Earth at its "center", is
discussed. An upper limit ($10^{112}$),
and a lower limit ($10^{102}$), for ${\cal S}$ are introduced.
It is suggested that intermediate-mass black holes (IMBHs)
constitute all dark matter, and that they dominate ${\cal S}$.
\end{abstract}


\bodymatter

\section{Introduction}

\bigskip

\noindent
Two references useful for further information about
the material of this talk are:

\noindent (1) P.H.F. and T.W. Kephart.
 {\it Upper and Lower Bounds on Gravitational Entropy}.
\noindent JCAP 06:008 (2008) and

\noindent (2) P.H.F.  {\it Identification of All Dark Matter as Black Holes}.
 {\tt arXiv:0905.3632 [hep-th]}.   JCAP 0910:016 (2009).

\section{The Entropy of the Universe.}

\bigskip

\noindent
As interest grows in pursuing alternatives to the Big Bang,
including cyclic cosmologies, it becomes more pertinent to
address the difficult question of what is the present entropy
 of the universe?

\bigskip

\noindent
Entropy is particularly relevant to cyclicity because it
does not naturally cycle but has the propensity only
to increase monotonically. In one recent proposal,
the entropy is jettisoned at turnaround. In any case, for cyclicity to be possible
there must be a gigantic reduction in entropy
(presumably without violation of the second law of thermodynamics)
of the visible universe at some time during each cycle.

\bigskip

\noindent
Standard treatises on cosmology
address the question of the  entropy of the
universe and arrive at a generic formula for a thermalized
gas of the form

\begin{equation}
S = \frac{2 \pi^2}{45} g_{*} V_U T^3
\label{thermal}
\end{equation}

\noindent 
where $g_{*}$ is the number of degrees of freedom, $T$ is
the Kelvin temperature and $V_U$ is the volume of the visible universe.
From Eq.(\ref{thermal}) with
$T_{\gamma}=2.7^{0}$K and $T_{\nu} = T_{\gamma} (4/11)^{1/3} = 1.9^{0}$K we find the
entropy in CMB photons and neutrinos are roughly equal today

\begin{equation}
S_{\gamma}(t_0) \sim S_{\nu}(t_0) \sim10^{88}.
\label{rad}
\end{equation}

Our topic here is the gravitational entropy, $S_{grav} (t_0)$.
Following the same path
as in Eqs. (\ref{thermal},\ref{rad}) we obtain for a thermal
gas of gravitons $T_{grav} = 0.91^{0}$K and then

\begin{equation}
S_{grav}^{(thermal)}(t_0) \sim 10^{86}
\label{thermalgrav}
\end{equation}

\bigskip

This graviton gas entropy is a couple of orders of magnitude below that for
photons and neutrinos.
This graviton gas entropy is a couple of orders of magnitude below that for
photons and neutrinos.
But there are larger contributions to gravitational entropy from elsewhere!!!

\bigskip

\section{Upper Limit on the Gravitational Entropy.}

\bigskip

\noindent
We shall assume that dark energy has zero entropy
and we therefore concentrate on the gravitational entropy associated with dark matter.
The dark matter is clumped into halos with typical mass
$M(halo) \simeq 10^{11}M_{\odot}$ where
$M_{\odot} \simeq 10^{57} GeV \simeq 10^{30} kg$
is the solar mass and radius $R(halo) = 10^5 pc \simeq 3 \times 10^{18} km \simeq
10^{18} r_S(M_{\odot}$). There are, say, $10^{12}$
halos in the visible universe
whose total mass is $\simeq 10^{23} M_{\odot}$ and
-corresponding Schwarzschild radius is $r_S(10^{23}
M_{\odot}) \simeq 3 \times 10^{23} km\simeq  10 Gpc$.
This happens to be the radius of the visible
universe corresponding to the critical density. This
has led to an upper limit for the gravitational entropy
is for one black hole with mass $M_U = 10^{23} M_{\odot}$.

\noindent 
Using $S_{BH}(\eta M_{\odot}) \simeq 10^{77} \eta^2$ corresponds to the
holographic principle for the upper limit
on the gravitational entropy of the visible universe:

\bigskip

\begin{equation}
S_{grav} (t_0)
\leq  S_{grav}^{(HOLO)}(t_0) \simeq 10^{123}
\label{HOLO}
\end{equation}.

\noindent
which is 37 orders of magnitude greater than
for the thermalized graviton gas in Eq.(\ref{thermalgrav})
and leads us to suspect (correctly) that Eq.(\ref{thermalgrav})
is a gross underestimate. Nevertheless, Eq.(\ref{HOLO})
does provide a credible upper limit, an overestimate yet to be refined downwards below,
on the quantity of interest, $S_{grav}(t_0)$.

\bigskip

\noindent
The reason why a thermalized gas of gravitons grossly underestimates
the gravitational entropy is because of the 'clumping'
effect on entropy.
Because gravity is universally attractive its entropy
is increased by clumping. This is somewhat counter-intuitive
since the opposite is true for the familiar 'ideal gas'. It is
best illustrated by the fact that a black hole always has
'maximal' entropy by virtue of the holographic principle.

\bigskip

\section{Lower Limit on Gravitational Entropy}

\bigskip

\noindent
It is widely believed that most, if not all, galaxies
contain at their core a supermassive black hole with
mass in the range $10^5 M_{\odot}$ to $10^9 M_{\odot}$
with an average mass about $10^7 m_{\odot}$. Each of these
carries an entropy
$S_{BH} ({\rm supermassive}) \simeq 10^{91}.$
Since there are $10^{12}$ halos this provides the lower limit
on the gravitational entropy  of

{\begin{equation}
S_{grav} (t_0) \geq 10^{103}
\label{lowerlimit}
\end{equation}

\noindent 
which, by now. provides an
eight
order of magnitude window for $S_{grav} (t_0)$.

\noindent
The lower limit in Eq.(\ref{lowerlimit}) from the
galactic supermassive black holes may be largest contributor
to the entropy of the present universe but this seems to us
highly unlikely because they are so very small. Each supermassive
black hole is about the size of our solar system or smaller
and it is intuitively unlikely that essentially
all of the entropy is so concentrated.

\bigskip

\noindent
Gravitational entropy is associated with the clumping
of matter because of the long range unscreened nature
of the gravitational force. This is why we propose that
the majority of the entropy is associated with the
largest clumps of matter: the dark matter halos associated
with galaxies and cluster.

\bigskip

\section{Intermediate Mass Black Holes}

\bigskip

\noindent
If we consider normal baryonic matter, other than black holes,
contributions to the entropy are far smaller. The background
radiation and relic neutrinos each provide $\sim 10^{88}$.
We have learned in the last decade about the dark side
of the universe. WMAP suggests that the pie slices
for the overall energy are 4\% baryonic matter, 24\% dark matter
and 72\% dark energy. Dark energy has no known microstructure,
and especially if it is characterized only by a cosmological
constant, may be assumed to have zero entropy. As already
mentioned, the baryonic matter other than the SMBHs contributes
far less than $(S_U)^{min}$.

\bigskip

\noindent
This leaves the dark matter which is concentrated in halos
of galaxies and clusters.

\bigskip

\noindent
It is counter to the second law of thermodynamics
when higher entropy states are available
that essentially all the entropy
of the universe is concentrated in SMBHs. The Schwarzschild
radius for a $10^7 M_{\odot}$ SMBH is $\sim 3 \times 10^7$ km
and so $10^{12}$ of them occupy only $\sim 10^{-36}$
of the volume of the visible universe.

\bigskip

\noindent
Several years ago important work by Xu and Ostriker
showed by numerical simulations that IMBHs with masses above
$10^6 M_{\odot}$ would have the property of disrupting
the dynamics of a galactic halo leading to
runaway spiral into the center. This provides
an upper limit $(M_{IMBH})^{max} \sim 10^6 M_{\odot}$.

\bigskip

\noindent 
Gravitational lensing observations are amongst the most useful
for determining the mass distributions of dark matter. Weak
lensing by, for example, the HST shows the strong distortion
of radiation from more distant galaxies by the mass of the
dark matter and leads to astonishing three-dimensional maps
of the dark matter trapped within clusters. At the scales
we consider $\sim 3 \times 10^7$ km, however, weak lensing
has no realistic possibility of detecting IMBHs in the
forseeable future.

\bigskip

\noindent 
Gravitational microlensing presents a much more optimistic
possibility. This technique which exploits the amplification
of a distant source was first emphasized in modern times
(Einstein considered it in 1912 unpublished
work) by Paczynski. Subsequent observations
found many examples of MACHOs, yet insufficient
to account for all of the halo by an order of magnitude.
These MACHO searches looked for masses in the range
$10^{-6} M_{\odot}
\leq M \leq 10^2 M_{\odot}$.

\bigskip

\noindent 
The time $t_0$ of a microlensing
event is given by

\bigskip

\begin{equation}
t_0 \equiv \frac{r_E}{v}
\label{t0}
\end{equation}

\bigskip

\noindent 
where $r_E$ is the Einstein radius and $v$ is the
lens velocity usually taken as $v = 200$ km/s. The radius
$r_E$ is proportional to the square root of the lens mass
and numerically one finds

\bigskip

\begin{equation}
t_0 \simeq 0.2 y ~ \left( \frac{M}{M_{\odot}} \right)^{1/2}
\label{t0value}
\end{equation}

\bigskip

\noindent
so that,  for the MACHO masses considered, $2 h \leq t_0
\leq 2 y$.

\section{ Cosmological Entropy Considerations.}

\bigskip

\noindent 
The cosmological entropy range

\bigskip

\noindent 
\begin{equation}
102 \leq \log_{10} S_U \leq 112
\label{entropy}
\end{equation}

\bigskip

\noindent 
is the first of two interesting
windows
which are the subject. Conventional wisdom is
$S_U \sim (S_U)^{min} = 10^{102}$.

\bigskip

\section{Intermediate Mass Black Holes and Microlensing Longevity}

\bigskip

\begin{center}

\begin{tabular}{||c|c|c|c||c||}
\hline\hline
$\log_{10} n_{max}$  & $ \log_{10} \eta$  & $ \log_{10} S_{halo}$ & $\log_{10}
S_U$ & $t_0$ (years) \\
\hline\hline
8 & 2 & 88 & 100 & 2   \\
\hline
7 & 3 & 89 & 101 & 6 \\
\hline
6  & 4 & 90 & 102 & 20 \\
\hline
5 & 5 & 91 & 103 & 60 \\
\hline
4 & 6 & 92 & 104 & 200 \\
\hline\hline
\end{tabular}

\bigskip

\noindent (Assumes $\rho_{IMBH} \sim 1\% \rho_{DM}$)

\end{center}

\bigskip

\section{ Observation of IMBHs}

\bigskip

\noindent
Since microlensing observations
already impinge on the lower end of the range
(\ref{DMBHmass}) and the Table, it is likely that observations
which look at longer time periods, have
higher statistics or sensitivity to the
period of maximum
amplification can detect heavier
mass IMBHs in the halo.
If this can be achieved, and it seems a worthwhile
enterprise, then the known entropy of the universe
could be increased by more than two orders of
magnitude.

\bigskip

\noindent
There exists interesting other analyses pertinent to
existence of massive halo objects:

J. Yoo, J, Chanam\'{e} and A. Gould, Astrophys. J. {\bf 601,} 311 (2004).
{\tt astro-ph/0307437}.

\bigskip

\noindent
It is this entropy argument based on holography and the second law of
thermodynamics which is the most
compelling supportive argument for IMBHs.
If each galaxy halo asymptotes to a black hole the final entropy
of the universe will be $\sim 10^{112}$
as in Eq.(\ref{entropy}) and the universe will
contain just $\sim 10^{12}$ supergigantic black holes.
Conventional wisdom is that the present entropy due
entirely to SMBHs is only $\sim 10^{-10}$ of
this asymptopic value.
IMBHs can increase the fraction
up to $\sim 10^{-8}$, closer to asymptopia
and therefore more probable according to the second law of thermodynamics.

\bigskip

\noindent
There are several previous arguments
about the existence of IMBHs and they have put upper limits
on their fraction of the halo mass.
The entropy arguments are new and provide additional motivation
to tighten these upper bounds or discover the halo black holes.
One observational method
is high longevity microlensing events. It is up to the
ingenuity of observers to identify other,
possibly more fruitful, methods some of which have
already been explored in a preliminary way.

\bigskip

\section{Post-conference update}

\bigskip

\noindent
Since the SCGT09 conference took place, the paper:

\bigskip

\noindent
Paul H. Frampton, Masihiro Kawasaki,
Fuminobu Takahashi and Tsutomu Yanagida

\noindent IPMU-09-0157 (December 2009).

\noindent {\it Primordial Black Holes as All Dark Matter}

\noindent {\tt arXiv:1001.2308 [hep-ph]}.

\bigskip

\noindent 
shows that it is possible to form black holes in the
early universe with mass $10^5 M_{\odot}$ and with
sufficient abundance to provide all of the dark matter.

\bigskip
\bigskip

\section{Acknowledgements}

\bigskip

\noindent
This work was supported in part by
the World Premier International Research Center Initiative 
(WPI initiative), MEXT, Japan and by U.S. 
Department of Energy Grant No. DE-FG02-05ER41418. 

\end{document}